\documentclass[twocolumn,showpacs,preprintnumbers,amsmath,amssymb]{revtex4}
\usepackage{graphicx}
\usepackage{dcolumn}
\usepackage{bm}

\newcommand{\vect}[1]{\bm{#1}}

\newcommand{\Fig}[1]{Fig.\ref{#1}}
\newcommand{\deffig}[4]{
  \begin{figure}
  \includegraphics[scale=#3]{#2}
  \caption{\label{#1} #4}
  \end{figure}
}

\begin{document}

\title{
Quadrupolar Order in Isotropic Heisenberg Models\\
with Biquadratic Interaction
}
\author{Kenji~Harada}
\affiliation{
Department of Applied Analysis and Complex Dynamical Systems, 
Kyoto University, Kyoto 606-8501, Japan}
\author{Naoki~Kawashima}
\affiliation{
Department of Physics, Tokyo Metropolitan University, 
Tokyo 192-0397, Japan}

\date{\today}

\begin{abstract}
Through Quantum Monte Carlo simulation, 
we study the biquadratic-interaction model with
the SU(2) symmetry in two and three dimensions.
The zero-temperature phase diagrams for the two cases
are identical and exhibit an intermediate phase 
characterized by finite quadrupole moment,
in agreement with mean-field type arguments and 
the semi-classical theory.
In three dimensions, we demonstrate that the model
in the quadrupolar regime has a phase transition
at a finite temperature.
In contrast to predictions by mean-field theories,
the phase transition to the quadrupolar phase 
turns out to be of the second order.
We also examine the critical behavior in the two
marginal cases with the SU(3) symmetry.
\end{abstract}

%
%

\maketitle

Spin interactions of the order higher than second
has been discussed for many years\cite{Suzuki1969,BlumeH1969}.
There are various sources of the high order terms.
For example, they may arise from the effect of crystalline fields,
or the high order perturbations of electron exchanges.
These high order terms were identified or speculated to be
responsible for some of the phase transitions observed 
in various magnetic materials
\cite{Kanamori1960,ElliottYS1971,SettaiETAL1998,Morin1988,Ohkawa1983,ShiinaST1997,MorinSTL1982}.
In contrast to the second order or bilinear interaction models,
quantum spin models with the high order terms
can have a phase diagram qualitatively different
from their classical counterparts.
In particular, at zero temperature they may have non-magnetic ordered phases
such as quadrupolar phase phase.
These non-magnetic phases have been a focus of attention
in recent years\cite{ShiinaST1997}.

In order for a higher order term to have a non-trivial
contribution to the Hamiltonian, the spin must be larger than
or equal to unity.
Among the simplest $S=1$ cases, we consider the model with 
the highest symmetry since it probably 
provides us with a good starting point for developing
a complete study of wider range of models with lower symmetry.
In the present article, therefore,
we discuss $S=1$ isotropic bilinear-biquadratic Heisenberg model:
\begin{equation}
  H = -\sum_{(ij)} \left( 
  J_L \vect{S}_i\cdot\vect{S}_j + J_Q (\vect{S}_i\cdot\vect{S}_j)^2 \right).
\label{eq:TheModel}
\end{equation}
Since the biquadratic term in this Hamiltonian arises from the fourth
order perturbation of electron exchanges,
it is usually smaller than the bilinear term that results from the
second order perturbation.
However, it was pointed out\cite{MilaZ2000} that
the bilinear term can be comparable with or smaller than the biquadratic one
as a result of cancellation of ferromagnetic and antiferromagnetic
contributions, when we take various hopping terms into account.

For the one-dimensional case, a number of exact solutions
and high-precision numerical calculations
have established the character of most of phases 
and the transition points.
For the two- or higher-dimensional cases, on the other hand,
our understanding largely depends upon mean-field type approaches 
or semi-classical theories\cite{Papanicolaou1986}.
A phase transition to non-magnetic ordered phase 
was predicted for a wide range of biquadratic models
including the present model.
The mean-field approximation\cite{ChenL1973}
was applied to the present model resulting that there is a 
first-order phase transition 
from the paramagnetic phase to the quadrupolar phase 
(or the spin-nematic phase)
when the biquadratic interaction is sufficiently large.

Since the mean-field type approaches are usually accompanied 
by uncontrollable errors, the confirmation through rigorous proof or 
numerical calculations is necessary.
In the classical model ($S=\infty$) with small $J_L/J_Q$,
it was rigorously proved
\cite{TanakaI1998} that the quadrupole moment is finite
in some temperature range above the dipolar transition point.
In the quantum case of $S=1$, 
the quadrupole moment was proved\cite{TanakaTI2001} to be 
finite at zero-temperature in some range of the parameter 
$J_L/J_Q$ in three dimensions.
The range where this rigorous proof applies
is not the same as, but smaller than the the quadrupolar region 
predicted by the mean-field arguments.
In two dimensions, there is no rigorous proof of existence of
the quadrupolar phase.

We reported in the previous work\cite{HaradaK2001} that
the parameter space of positive $J_Q$ is divided
by the two SU(3) points into three regions;
ferromagnetic ($-\pi \le \theta \le -3\pi/4$),
antiferromagnetic ($-\pi/2 \le \theta \le 0$),
and non-magnetic ($-3\pi/4 < \theta < -\pi/2$) regions.
Here $\theta$ is defined by
$$
  J_L = -J\cos\theta, \quad J_Q = -J\sin\theta \qquad (J >0).
$$
The nature of the non-magnetic phase was not numerically
identified in the previous work,
although the mean-field theory predicted that it is 
the quadrupolar phase.
In the present paper, we show for the model with negative $\theta$
that (1) the non-magnetic phase is characterized by the finite quadrupole
moment in two and three dimensions, 
(2) a phase transition to quadrupolar phase occurs at a finite temperature
in three dimensions, 
and 
(3) the quadrupolar transition is of the second order in contrast to 
the mean-field prediction.
We also discuss the critical behavior of the three-dimensional system
at finite temperature.

In the classical counterpart of the present model,
the long range order at zero temperature is 
always dipolar, i.e., ferromagnetic or antiferromagnetic,
except for the special case of $\theta = -\pi/2$,
where the dipolar degrees of freedom are non-interacting and disordered.
In contrast, for the quantum model for $S=1$,
it is argued based on a mean-field approximation\cite{ChenL1973} 
that there is an intermediate phase between the anriferromagnetic
region and the ferromagnetic region,
and that this phase is characterized by finite quadrupole moment.
Because of the limitation of the mean-field-type theory,
it always predicts a finite temperature phase transition to
the quadrupolar phase regardless of the dimensionality.
This is of course wrong in one dimension.
In two dimensions, too, the existence of finite temperature phase
transition is very questionable because of 
the Mermin-Wagner theorem.
Even at zero temperature, the existence of the finite quadrupole
moment is not totally clear.
Mathematically rigorous arguments, so far, has not established
any long-range order in the intermediate parameter region.

In order to answer to the question concerning
the existence of the quadrupole order,
we performed Monte Carlo simulation using the loop algorithm proposed 
in the previous letter\cite{HaradaK2001}.
The algorithm removes the ergodicity problem and
considerably reduces the critical slowing down.
The energy $E$, the dipole moment (i.e. magnetization) $M_z$,
the staggered magnetization $N_z$, and
the quadrupole moment, were measured.
We consider only the $zz$ component of the quadrupole moment
in this article, which we denote by $Q_z$;
$$
  Q_z \equiv \sum_i \left((S^z_i)^2 - \frac23\right).
$$
The equal-time structure factors and the susceptibilities 
associated to these quantities were also measured.
The system size ranges from $L=4$ up to $L=128$ for two-dimensional
case and up to $L=64$ for three-dimensional case.
For each data point, we typically run the simulation 
for more than $10000$ Monte Carlo Steps.

For each system size in two dimensions, the thermal average of the
absolute value of the quadrupole moment, $q \equiv \langle |Q_z|
\rangle/N$ converges to a certain finite value as the inverse
temperature $K\equiv J/k_{\rm B}T$ increases.  
Here, the absolute value $|Q_z|$ is taken in the representation basis
in which $Q_z$ is diagonalized.
For any finite system, the convergence
is exponential with some characteristic (imaginary) time scale.
Although this characteristic time is larger for larger systems, the
size dependence is weak.  Therefore, we can extrapolate the data to
the limit of $K=\infty$ without examining extremely low
temperatures.  After taking the zero-temperature limit numerically, we
then take the infinite system size limit.  The system size dependence
is algebraic;
$$
  q(L,K=\infty) \sim q(L=\infty,K=\infty) + a/L.
$$
This system size dependence is the same as that of
the staggered magnetization in the antiferromagnetic Heisenberg
model in two dimensions.

Quadrupole moment at zero temperature as a function of $\theta$
for various system sizes is plotted in \Fig{fig:2d_aqz_vs_THETA},
together with the extrapolation to infinite size.
\deffig{fig:2d_aqz_vs_THETA}{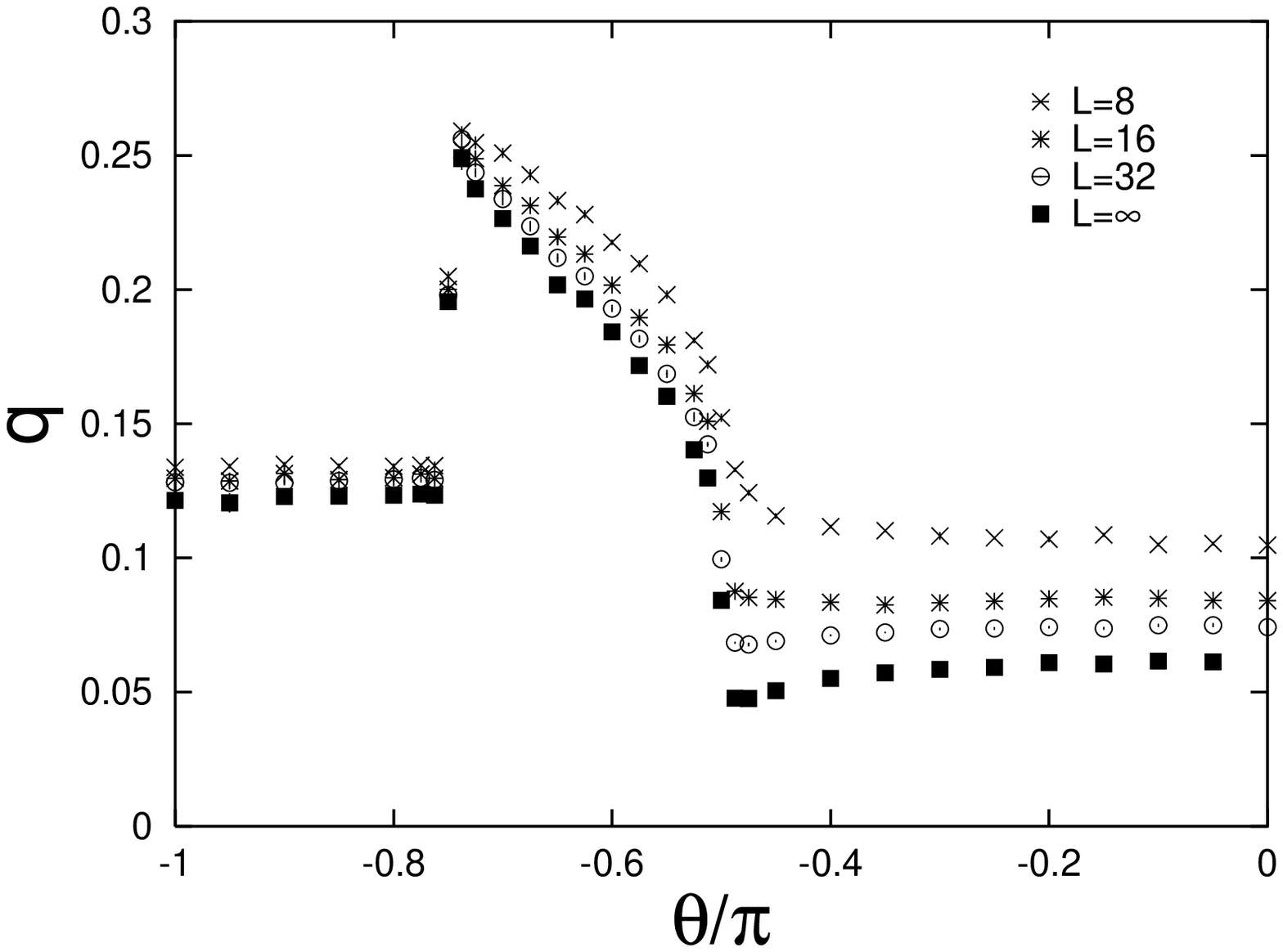}{0.45}
{The quadrupole moment at zero temperature in two dimensions.}
We now see that the quadrupole moment is finite in the
intermediate phase as well as in the dipolar phases.
In addition, it exhibits discontinuity at the two SU(3) points.
Since the quadrupole moment is finite 
whenever the dipole moment is finite, 
it falls down to a finite value, not to zero, as we 
pass the phase boundary from the intermediate region to one of 
the two dipolar regions.
Since the dipole moment is vanishing in the intermediate phase
as we saw in the previous paper\cite{HaradaK2001},
the quadrupole moment is the characterizing order parameter
for this phase.

In order to check the existence or absence of a phase transition
at a finite temperature, we have examined the specific heat.
We have observed a broad peak at the temperature
that roughly corresponds to the saturation temperature
of the quadrupole moment.
The peak height and width do not show a significant size dependence, 
indicating that it is not a phase transition but a point where
a gradual change from the paramagnetic state to the
quadrupolar state takes place.

We plot in \Fig{fig:2d0600_aqz_vs_L}
the size dependence of the quadrupole moment
as a function of system size, at various temperature
in the case of $\theta = -0.6\pi$.
In this case, the peak in the specific heat is 
located at $K \sim 1.2$.
\deffig{fig:2d0600_aqz_vs_L}{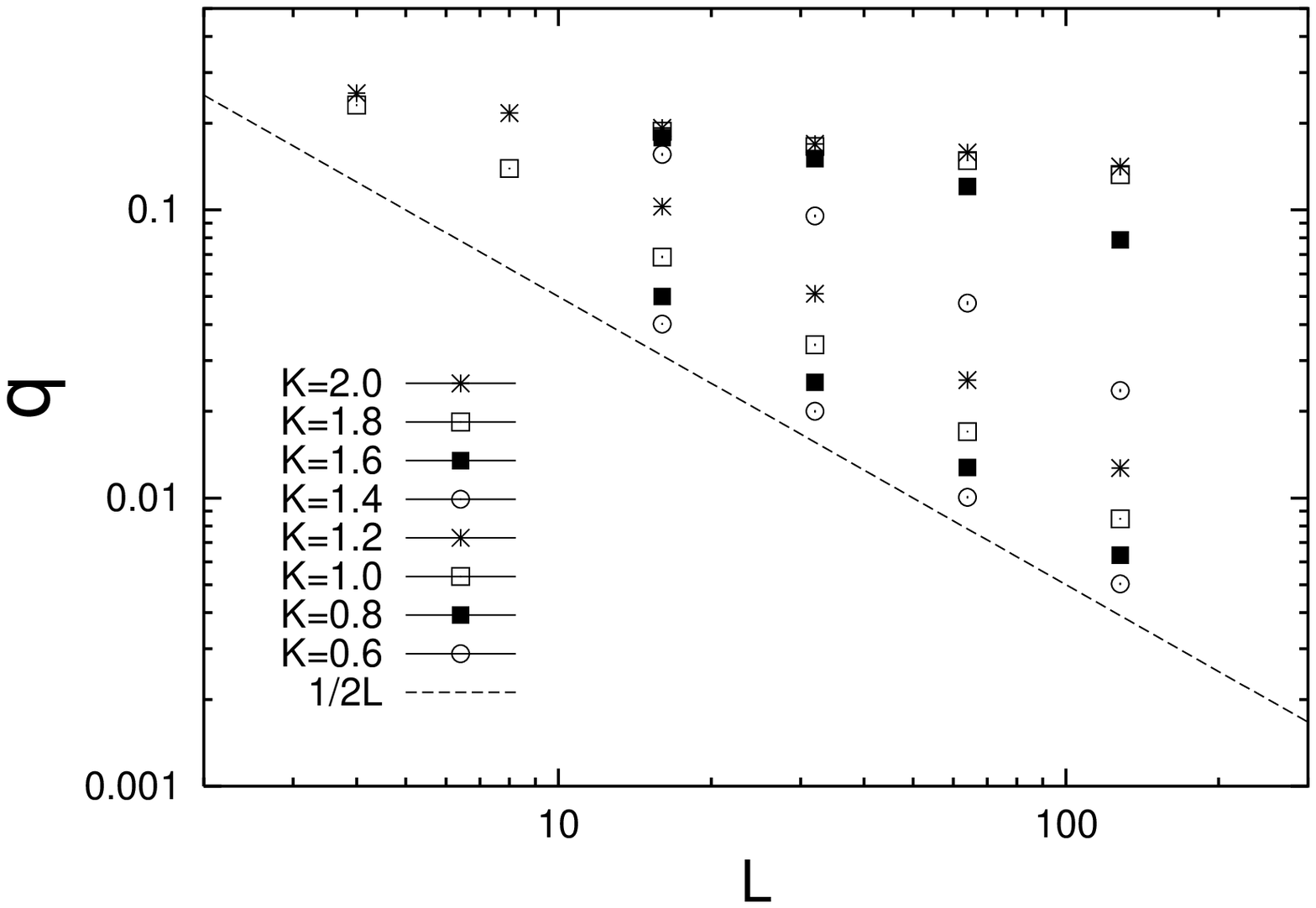}{0.45}
{The quadrupole moment as a function of the system size at
$\theta = -0.6\pi$ in two dimensions.}
In \Fig{fig:2d0600_aqz_vs_L},
we see that the quadrupole moment shows the asymptotic size dependence
$$
  q \propto L^{-1}
$$
down to the temperature $K \sim 1.6$.
For the temperature lower than $1.6$, the largest system size that
we examined is not large enough to see the asymptotic behavior.
The transition temperature of the KT-type phase transition
is usually about 10 or 20 percent smaller than the peak 
temperature of the specific heat.
Therefore, if there were a Kosterlitz-Thouless type transition,
we should be able to see a non-trivial algebraic decay for $K < 1.6$,
which we do not detect in \Fig{fig:2d0600_aqz_vs_L}.
This indicates that there is no phase transition
at any finite temperature.

For the system in three dimensions at zero temperature, 
we again observe three parameter regimes;
ferromagnetic, quadrupolar, and antiferromagnetic,
with exactly the same phase boundaries 
as those in two dimensions.
Namely, the nature of the ground state changes at the
two SU(3) points, $\theta = -\pi/2, -3\pi/4$.
To see this in detail, we analyze the order parameters
as in the two-dimensional case;
the extrapolation to zero temperature,
and then to the infinite system size.
The behavior of the zero-temperature quadrupole moment as a function
of $\theta$ is similar to the two-dimensional case, but
the convergence to the infinite size limit is faster.
The quadrupole moment shows discontinuity
at the two symmetric points.
The zero-temperature phase diagram in three dimensions
turns out to be exactly the same as that in two dimensions.
We speculate that this is true for any dimensions except for 
one dimension.

Having seen the long range order at zero temperature
in the intermediate quadrupolar regime,
we now ask if there is a phase transition at a finite temperature.
Even in two dimensions, we have seen a broad peak in the
specific heat and a cross-over behavior from
completely disordered states to partially ordered states
as we decrease the temperature.
This may be regarded as a precursor to the phase transition
in higher dimensions.
In fact, in the specific heat as a function of the temperature,
we see a much sharper peak in three dimensions
than in two dimensions.
The peak is not only sharp but also shows clear size dependence,
indicating a phase transition.

\deffig{fig:3d0600_sen_FSS}{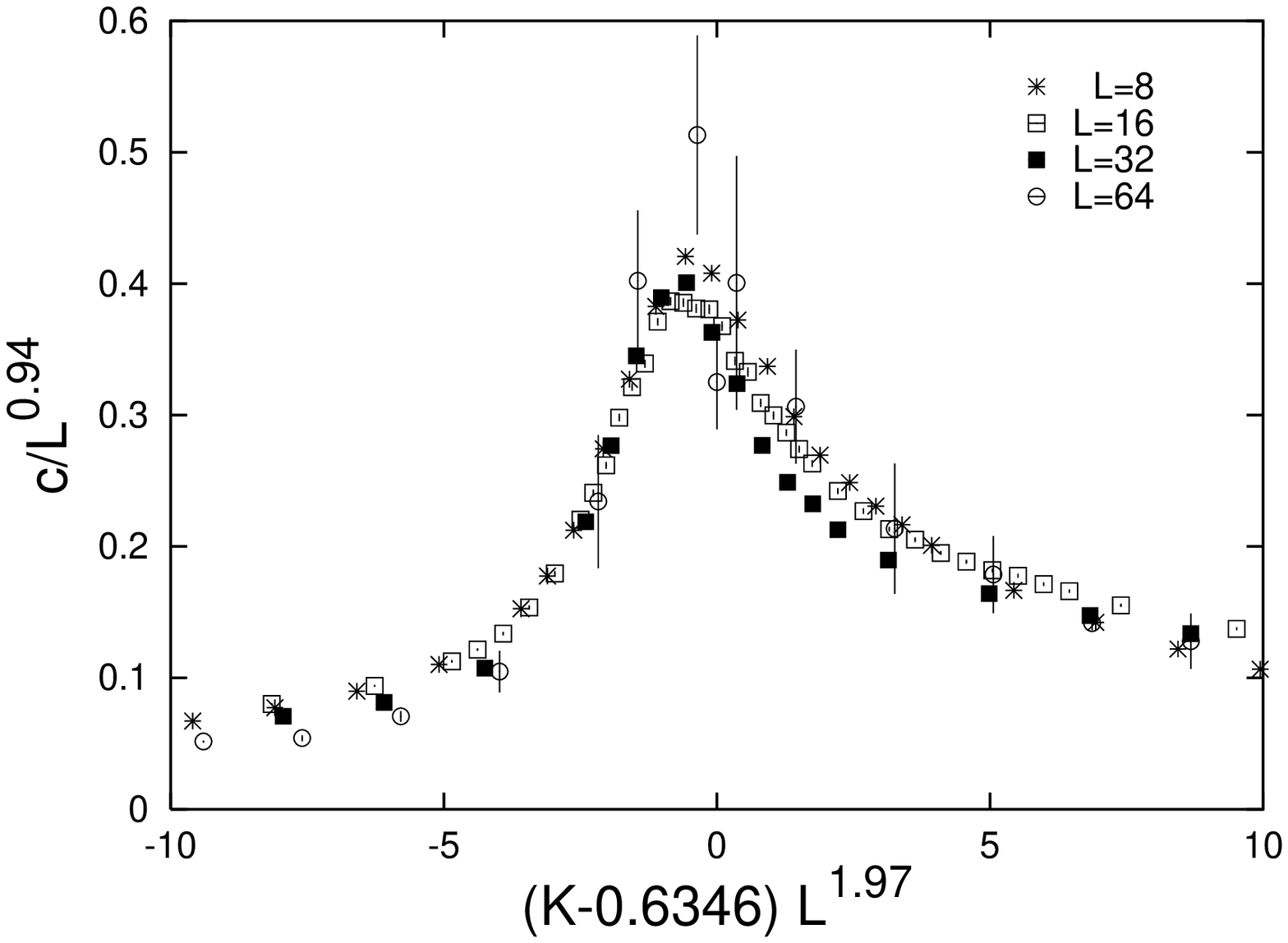}{0.45}
{Scaling plot of the specific heat at $\theta=-0.6\pi$ in three dimensions}
\deffig{fig:3d0600_aqz_FSS}{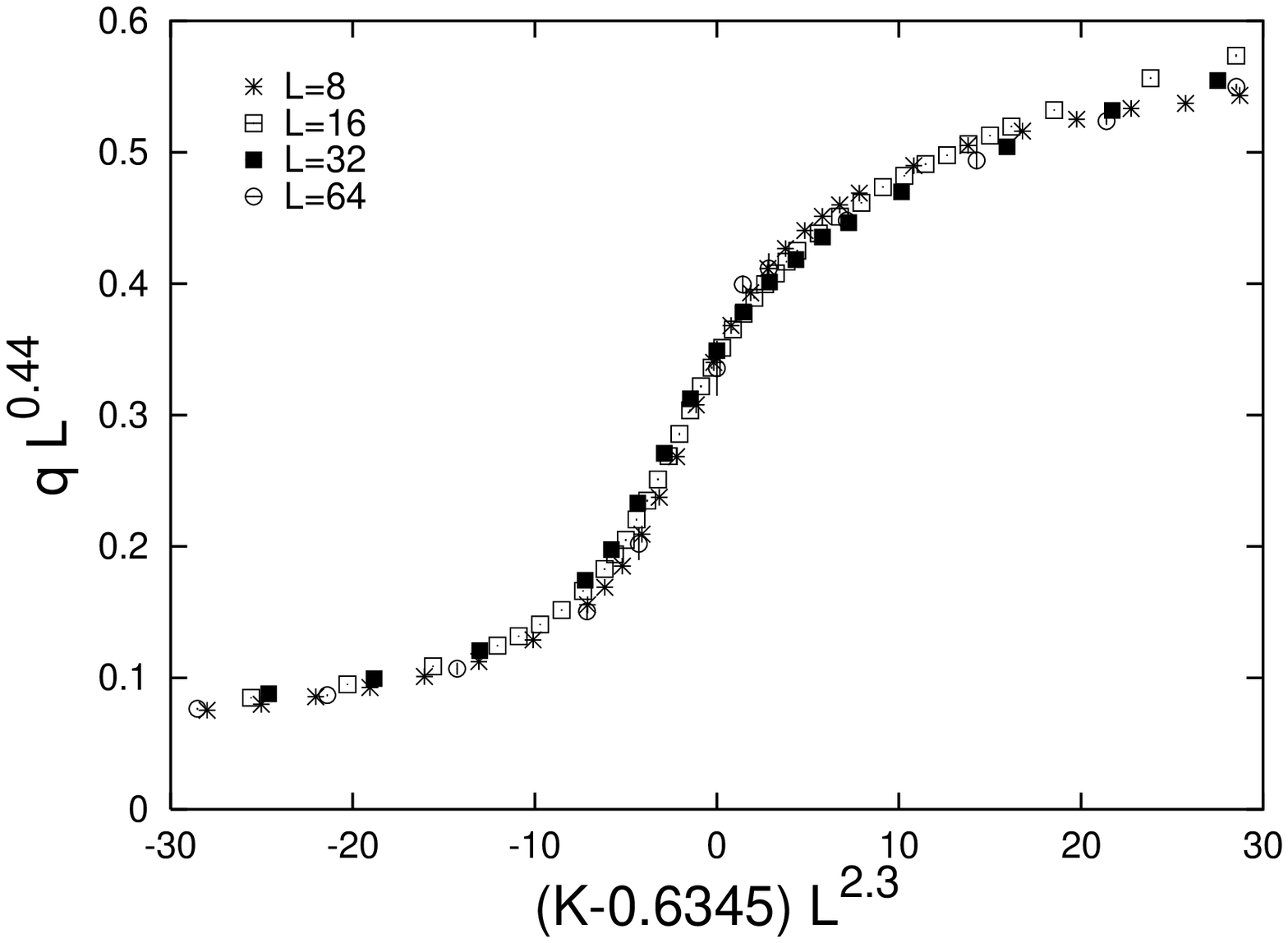}{0.42}
{Scaling plot of the quadrupole moment at $\theta=-0.6\pi$ in three dimensions}
We can clearly see a strong correction to scaling, especially in 
\Fig{fig:3d0600_sen_FSS}.
According to the mean-field theory\cite{ChenL1973}, this is
a first order phase transition.
If this is the case, the peak height and width should be
proportional to $L^d$ and $L^{-d}$, respectively.
In other words, a finite size scaling plot with exponents 
$d$ and $-d$ for the vertical scale and the horizontal scale, 
respectively, should work.
Other quantities should also obey similar scaling forms 
with trivial exponents.
We find that this is obviously not the case for any quantity.
Instead, we assume the following finite-size-scaling forms;
\begin{eqnarray*}
  c & = & L^{2y-d}\tilde c((K-K_c)L^y), \\
  q & = & L^{\beta y}\tilde q((K-K_c)L^y).
\end{eqnarray*}
where $c$ is the specific heat.
The best plots are obtained with
$$
  y = 1.97\quad\mbox{and}\quad K_c = 0.6346,
$$
for the specific heat, and
$$
  y = 2.30,\quad y\beta = 0.44 \quad \mbox{and} \quad K_c = 0.6345,
$$
for the quadrupole moment.
The scaling plots are shown 
in \Fig{fig:3d0600_sen_FSS} for the specific heat 
and in \Fig{fig:3d0600_aqz_FSS} for the quadrupole moment.
The discrepancy among the estimates of indices 
may be due to a relatively large contribution of
the non-singular part to the specific heat.
We have estimated the critical temperatures and indices at $\theta = -0.7\pi$
in a similar fashion, 
and found that the critical indices are close to the corresponding
ones for $\theta = -0.6\pi$ quoted above.
This fact suggests that they belong to the same universality class, 
as expected.
Based on these results, we conclude
$$
  y = 2.15(20), \quad \mbox{and} \quad y\beta = 0.46(4)
$$
for $-3\pi/4 < \theta < -\pi/2$.

The two SU(3) points, $\theta = -\pi/2$ and $-3\pi/4$, 
are of special interest, since the universality
class of the critical point may be different from the one discussed
above due to the higher symmetry.
For these points of higher symmetry,
we obtained better scaling plots
than \Fig{fig:3d0600_sen_FSS} and \Fig{fig:3d0600_aqz_FSS}.
The estimated critical temperatures are
$K_c = 0.6389(3)$ for $\theta = -\pi/2$ and 
$K_c = 1.0724(4)$ for $\theta = -3\pi/4$.
The critical indices for these two cases agree with each other, yielding
$$
 y = 1.82(5), \quad y\beta = 0.48(1).
$$
These result suggest that the critical points of the two SU(3) 
models belong to the same universality class and 
it is distinct from the one for the less symmetric cases
although the difference in the indices is small.
In \Fig{fig:PhaseDiagram}, we summarize the estimated critical 
temperatures in the form of a $\theta$-$T$ phase diagram.
\deffig{fig:PhaseDiagram}{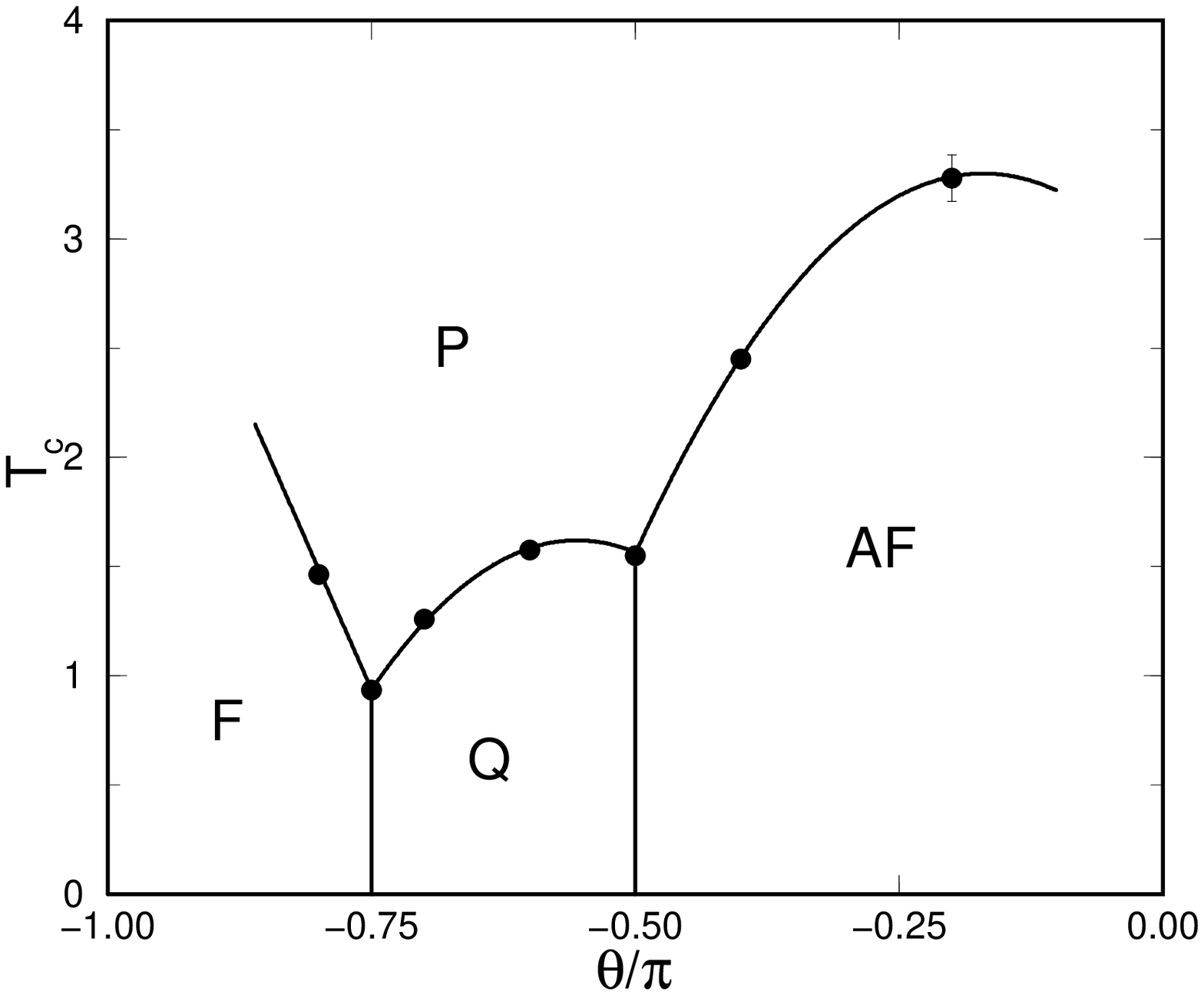}{0.45}
{The phase diagram of the three-dimensional model.
Curves are mere guides to eyes.}

To summarize, we have studied the isotropic biquadratic Heisenberg model
in two and three dimensions for negative $\theta$.
In two dimensions, we have identified the intermediate phase as
the quadrupolar phase.
The phase transition at a finite temperature has been excluded.
In three dimensions, we have studied finite temperature properties
as well as zero temperature ones.
At zero temperature, the phase diagram is exactly the same as that
of the two-dimensional case.
We have found that there is a finite temperature phase transition
not only in the ferro- and antiferro- magnetic
regimes but also in the quadrupolar regime.
In contrast to the mean-field prediction, the transition to 
the quadrupolar phase has turned out to be of the second order.
The critical indices are also estimated.
While the two SU(3) symmetric points belong to the same universality
class, it is suggested to be distinct from the one for the less
symmetric (i.e., SU(2)) models.
Studies on the properties of low-lying excitations
are still in progress and will be reported elsewhere.
Less symmetric models with higher order interactions
may be more important than the present model from the practical point of view, 
since the higher order interactions in real magnets
often arise from the crystalline effects, which have lower symmetry.
Studies on some of these models are also in progress.


The authors thank C.~Batista, G.~Ortiz, \hbox{J.~E.~Gubernatis} 
and Y.~Okabe
for their useful comments.
A part of N.K.'s work was done while he was staying at University of
Cergy Pontoise, France.
He is grateful to H.-T.~Diep for his hospitality.
The computation was performed on SGI Origin 2800/384 at 
Supercomputer Center, University of Tokyo,
Institute of Solid State Physics, University of Tokyo.
The present work is financially supported by
Grant-in-Aid for Scientific Research Programs
(No.11740232 and No.12740232) from JSPS, Japan.

\end{document}